# Automated Quality Assessment of (Citizen) Weather Stations

Today we have access to a vast amount of weather, air quality, noise or radioactivity data collected by individual around the globe. This volunteered geographic information often contains data of uncertain and of heterogeneous quality, in particular when compared to official in-situ measurements. This limits their application, as rigorous, work-intensive data cleaning has to be performed, which reduces the amount of data and cannot be performed in real-time. In this paper, we propose dynamically learning the quality of individual sensors by optimizing a weighted Gaussian process regression using an evolutionary algorithm. The evaluation is done for the south-west of Germany in August 2016 with temperature data from the Wunderground network and the Deutsche Wetter Dienst (DWD), in total 1,561 stations. Using a 10-fold cross-validation scheme based on the DWD ground truth, we can show significant improvements of the predicted sensor reading. In our experiment we were obtain a 12.5% improvement on the mean absolute error.

Julian Bruns[1], Johannes Riesterer[2], Bowen Wang[2], Till Riedel[2] and Michael Beigl[2]

[1] FZI Forschungszentrum Informatik, Germany

[2] Karlsruhe Institute of Technology, TECO / Pervasive Computing, Germany

***Keywords:*** Crowdsourcing air temperature; Data quality assessment; Evolutionary Learning; Gaussian process regression; Volunteered geographic information.

## Introduction

Today, we are living in an era where sensors are cheap, can be easily obtained and put into use with little effort – they become ubiquitous. In the field of geo-science in particular this leads to many new data sources and opportunities. In addition to classical data sources such as government organizations, individuals are now providing data voluntarily, so called volunteered geographic information (VGI). These information sources range from smartphones, GPS-equipped mobile devices to privately owned weather stations on such sites as Wunderground[1] or OpenSenseMap[2] to name but a few. Projects such as OpenStreetMap (OSM)[3] are empowering and encouraging individuals to provide data and participate in order to create an open map. All this possibilities could lead to a "collective [geo] sensing" (Blaschke et al. 2011). Even as of today, the increased availability of data sources leads to a highly increased resolution in both the spatial and temporal dimension.

---

[1] https://www.wunderground.com/

[2] https://opensensemap.org/

[3] https://www.openstreetmap.org/

Measurements can be done in-situ, at any given area of interest and can be re-located if the need arises.

But these new data sources come with new challenges regarding their use. To "produce results that can be trusted" (Stewart 2011), the quality and location of measurements has to be known. Traditional data sources are often standardized measurements of government agencies. Their quality and exact location is most of the times well-known, they are calibrated regularly and have almost 100% service time. VGI does not have this advantage. VGI are provided by different organizations and acquired differently. The resulting diversity in creditably, data structure etc. can add additional uncertainty to the result, which prevents their use without appropriate pre-processing. A good example is the recent study of Meier et al. (2017), in which they discussed the use of crowdsourced weather data for the city of Berlin in 2015. During their quality assessment they had to filter over 50% of the available data and stated that "rigorous data quality assessment is the key challenge" (Meier et al 2017). And while this quality assessment can be done by experts and on historical data, the associated workload is high. This is not feasible for "big data" or in real time.

The goal of this paper is to assess the quality of citizen science weather data to improve predictions and meteorological models using these new data sources. To solve their problem, we propose an automated quality assessment based on an evolutionary algorithm. Based on benchmark measurements the algorithm learns the quality of each sensor. We then apply the calibrated data in a Gaussian Process Regression (GPR) to predict the measurement of interest. Our approach allows us to incorporate expert knowledge as a-priori information in the evolutionary algorithm as well as iterative improvement of the quality assessment with each new measurement. It is derived from the field of ubiquitous computing as well as well-known approaches from spatial statistics. We evaluate the proposed approach with a temperature prediction for the area of south-west Germany and data from the Deutsche Wetterdienst and the Wunderground network. We use the equivalent to ordinary kriging as our basic Gaussian Process Regression to show the improvement even without additional background information.

# Related Work

## Crowd Sourced Sensing

In crowd sourced sensing, a group of private and/or professional users collect and contribute sensor information collaboratively to form a body of knowledge. Particularly the rise of smart phones and "the increasing ability to capture, classifying, and transmit a wide variety of data (image, audio, and location) have enabled a new sensing paradigm" [Reddy et al., 2007]. Civic agencies of several countries across the world are already harnessing the hive intelligence of the public by accelerating and scaling the use of such open innovation methods to help address a wide range of urban and societal problems ranging from wildlife observations to air quality sensing (obamawhitehouse 2014). Participatory geo sensing was successfully applied to study physical phenomena particularly in city contexts such as urban noise levels as an alternative to traditional environmental monitoring (D'Hondt et al., 2013).

In an Internet of Things anything can be measured using „a set of observations that reduce uncertainty where the result is expressed as a quantity" (Hubbard, 2007). This statistically motivated view on measurement partially contradicts classical view on measurement processes as standardized as DIN 1319 that has shaped much of the last century. Considering the sparse spatial and temporal solution of many measurements available today, anything that is better than guessing (on top of existing knowledge) can potentially contribute to a measurement even if not even considered a measurement by strict definitions. However, this (as also addressed in our work) requires algorithms to cope with "the problem of interrelationship between reliability of information sources, their number, and the reliability of fusion results"(Rogova et al., 2004).

Early research, focused on mostly managing distributed sensor on sensor webs like Intel's IrisNet (Gibbons et al., 2003) or Microsoft's SenseWeb (Grosky et al., 2007), have long become reality with broad availability of Devices like NetAtmo and has been attracted the attention of researchers particularly interested in higher resolution data (Chapman et al., 2017), (Meier et al., 2017). In this work as well is the data used to interpolate fine grained temperature distributions. However, today only little objective research on the quality of this volunteered geographic information using a large number of measurements exists.

## Prediction of environmental factors

The main advantage of VGI it to get more data and information about the environment to formulate and evaluate hypothesis and gain valuable insights into the environment. Data gained is used to train models to predict environmental factors such as temperature and pollution. The basis for all spatial prediction models is Tobler's First Law (Tobler, 1970), stating that "everything is related to everything else, but near things are more related than distant things". One of the most often used approaches to incorporate this law is Kriging (Krige, 1951), which was developed to estimate ore deposits but is since used for predictions for a manifold of spatial applications and has been modified to be more powerful and general. Hengl et al. (2012) uses a kriging approach to predict temperatures. They include a temporal component to predict the daily mean temperature in Croatia for area of $1 km^2$ with an accuracy of 2.4 °C by combining Modis satellite images with 57,282 ground measures of daily temperatures in 2008. In a follow-up work, Kilibarda et al. (2014) introduce an automated mapping framework for predictions of daily mean, minimum and maximum air temperatures using regression-kriging for a resolution of 1km with a root-mean-square error between 2 and 4°C.

Gräler et al. (2016) developed an R-package called gstat, which uses copulas to enable a spatio-temporal kriging. They show the power how their approach with a prediction of daily mean PM10 concentration in 2005 in Germany.

Another modification of the kriging approach can be found in Bhattacharjee et al. (2016). They propose a semantic kriging approach, where a high-resolution satellite snapshot is used to quantify the effect of the difference between different locations as well as the interaction of the land uses between those points. The different land use classes are learned in a semantic hierarchical network.

Hjort et al (2011) presented a different approach to predict local temperatures in the city of Turku, Finland. They used generalized linear models combined with regression trees and data from 36 stationary weather stations over a period of six years.

A fundamental overview and theoretical background as well as applications of spatio-temporal statistics is found in Cressie and Wikle (2015).

# Method

Our approach combines a novel evolutionary learning algorithms to automatically asses and determine the quality of each sensor and models this information as an uncertainty kernel. This is then combined with a typical ordinary kriging kernel as a GPR to predict temperature.

## Gaussian Process Regression

We want to modify a regression model such that it can take into account the individual quality of an observation. For this purpose the classical Kriging with noise is not suitable, since the noise factor can only model a constant additional quantity in contrast to the non-constant quality of the data points. It turned out that the more general Gaussian process regression meets our requirements since it is determined by defining a covariance function and we are therefore able to model the quality of measurements by constructing the appropriate covariance function. In particular we combine a Matern covariance function with a covariance function that maps a quality parameter of an observation to an uncertainty of its correctness. We use a Matern covariance function for the following reasons. Since every physical process is of local nature we may assume that the measurement of temperature on earth follows Tobler's first law of geography. Furthermore we may assume that local fluctuations still can occur due to meteorological and topographical effects. The limit of a Matern covariance function yields an exponential covariance function and thus realizes Tobler's first law of geography. However appropriate choices of the parameters result in less smooth functions which are more suitable to fit the mentioned fluctuations but are still smooth enough to be robust against statistical noise.

For the following paragraphs about Gaussian Processes, regressions and modelling compare Edward et al (2006), in particular Chapter 4 for definitions and properties of covariance functions.

Let y be the quantity we want to predict at a point p and $D = \{(p_i, y_i, q_i) | i \in [1 \dots n]\}$ be a set of data points, where $p_i$ denotes a specific point in geo-coordinates, $y_i$ an observation of y at

point $p_i$ and $q_i$ the quality parameter of the measurement. We assume that the observations are measurements of a physical process and thus they can be assumed to follow Tobler's first law of geography as already motivated above. If we furthermore assume that the errors of the measurements are normal distributed, it is reasonable by definition to model the quantity y as a Gaussian process.

We define the function

$$\kappa_Q(q_i, q_j) := \begin{cases} \frac{\lambda}{q_i^2} & \text{if } i = j \\ 0 & \text{else} \end{cases}$$

for the quality parameters of two observations, where $\lambda > 0$ is a fixed scaling parameter. It is a covariance function since it is positive everywhere and only nonzero on the diagonal. Furthermore let $\kappa_M(d(p_i, p_j))$ be a Matern covariance function with respect to the distance $d(p_i, p_j) = |p_i - p_j|$. Since the sum of two covariance functions is a covariance function,

$$\kappa\big((p_i, y_i, q_i), (p_j, y_j, q_j)\big) := \kappa_M(d(p_i, p_j)) + \kappa_Q(q_i, q_j)$$

also defines a covariance function.

For a subset $S \subset D$ we denote by $GPR_\kappa(p \mid S)$ the corresponding Gaussian process regression for the quantity y at a point p under the observation S, which is implemented by our new (combined) Kernel function: $\kappa\big((p_i, y_i, q_i), (p_j, y_j, q_j)\big)$.

## Evolutionary Algorithm

We use an evolutionary algorithm to train the quality parameter. The algorithm iteratively generates new variants of the set of data points with modified qualities. We evaluate the fitness of each variant by considering the error of predictions with the Gaussian process regression.

As defined in the last section let $D = \{(p_i, y_i, q_i) \mid i \in [1 \dots n]\}$ be a set of data points, $S \subset D$ a subset and $GPR_\kappa(p \mid S)$ the corresponding Gaussian process regression for the quantity y at point p under the observation S. For another subsets $S' \subset D$ we define the fitness function $\text{fit}(S' \mid S) := \sum_{s' \in S'}(s' - GPR_\kappa(s' \mid S))^2$, which measures the error between observations in S' and their prediction by the Gaussian process regression under the observations S.

Let DWD denote the dataset of the Deutsche Wetterdienst and WG the dataset of Wunderground. They contain tuples of the form $(p_i, y_i)$, where $p_i = (\text{lat}_i, \text{long}_i)$ are geo-coordinates and $y_i$ is the measured temperature at this point. To evaluate our model using a 10-fold cross validation scheme, we apply a test train split to DWD which yields the decomposition into $DWD_{\text{valid}}$ and $DWD_{\text{train}}$.

In the training process (1 - 5), we build and use an evolutionary algorithm with without crossover. For each generation, the $D_{\text{cur}}$ is divided into $D_{\text{pred}}$, $D_{\text{unchanged}}$ and $D_{\text{mut}}$ in a 0.3 : 0.5 : 0.2 proportion. $D_{\text{pred}}$ is chosen to contain 20% of the data points with the highest quality in $D_{\text{cur}}$. The remaining points in $D_{\text{cur}}$ are assigned randomly.

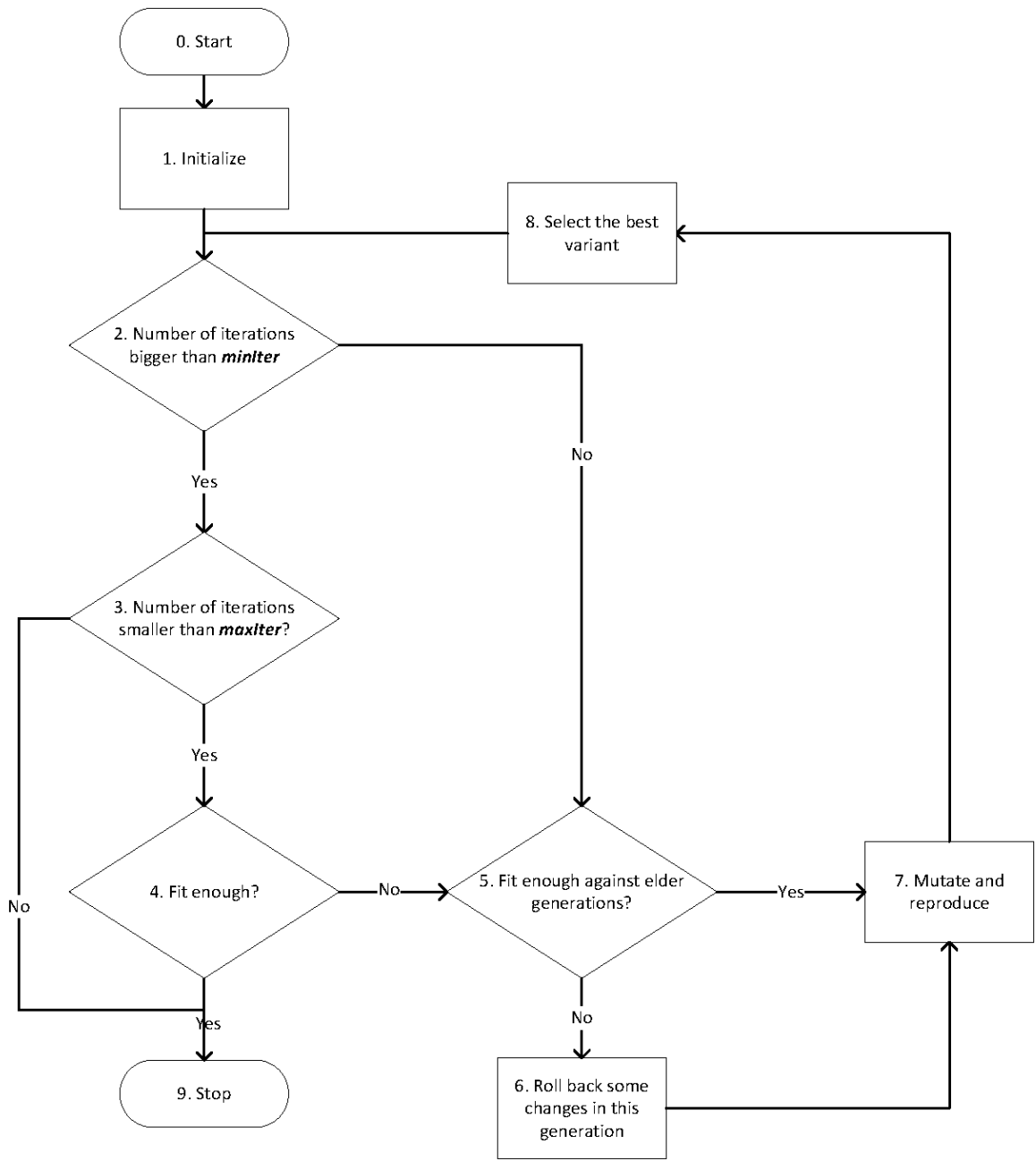

*Figure 1: Graphical description of the evolutionary algorithm. At each iteration the termination criteria are checked and if negative, the mutation and learning process is performed.*

Each generation is evaluated with a fitness function based on the value of the MSE of predicting $D_{pred}$ of the generation. The fitness value determines whether the observations in $D_{pred}$ can be better reproduced by the parent generation $D_{cur} \setminus D_{pred}$, or by the new generation.

Our algorithm performs the following steps, the numbering is analog to Figure 1:

1: The training process is initialized with the population $D_{cur} := DWD_{train} \cup WG$ as the union of $DWD_{train}$ with WG, where the qualities are set to 1 for datapoints of $DWD_{train}$ and to some fixed value µ ∈ (0,1] for datapoints of WG respectively.

2-4: After at a minimal number of iterations given by the hyper parameter ***minIter***, the training process can be terminated if the improvement of the last few iterations is below a threshold and seems to have been converged If the training process exceeds the maximal number of iterations ***maxIter***, the training process will be forcedly terminated. In this work, ***minIter*** is set to 20 and ***maxIter*** to 100.

7: From $D_{mut}$ two mutations are reproduced: $D_{mut}^1 := \{(p_i, y_i, (0.9 * q_i + 0.1)) | (p_i, y_i, q_i) \in D_{mut}\}$ and $D_{mut}^2 := \{(p_i, y_i, (0.9 * q_i)) | (p_i, y_i, q_i) \in D_{mut}\}$ are created by randomly raising or lowering the quality of the elements in $D_{mut}$.

5-6: The fitness value of the current population $D_{cur}$ is evaluated against that of its elder generations. If $D_{cur}$ results in a worse fitness score, the current generation will roll back towards the last generation.

8: The variant from $\{D_{mut}, D_{mut}^1, D_{mut}^2\}$ that results in the highest fitness score will be selected. We replace $D_{mut}$ with the selected variant $D_{selected}$ to create the next generation.

9: The result of the algorithm is a quality value for each sensor, which is then used in the Gaussian Process Regression with our new combined covariance function.

## Evaluation

### Dataset

Our dataset is based on temperature measurements, taken each day at 12:00 MET within the latitude longitude range of [47.5; 49.5; 7.5; 9.5] for all weather stations of the DWD and Wunderground station networks. The models are trained on the data from the 01. – 04.08.2016 and evaluated from 05. – 08.08.2016. 42,966 observations are used from 1,561 weather stations. 48 stations are from the DWD and 1,513 are from the Wunderground network.

We use a ten-fold cross-validation for the learning approach and predict temperatures at randomly chosen DWD weather stations, which are removed from the training data set.

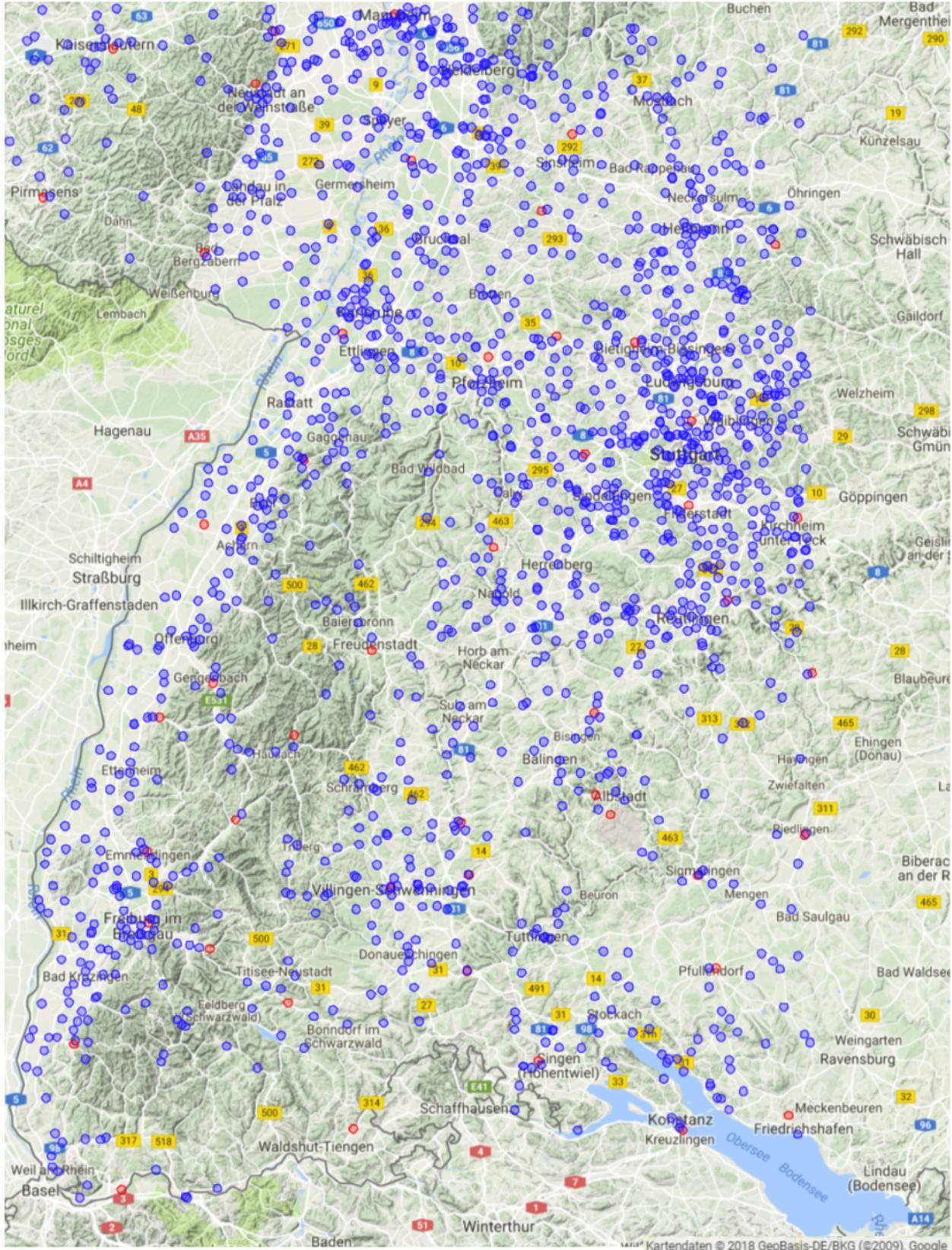

*Figure 2: Spatial distribution of the stations in Google Maps. In red are the DWD stations.*

## Parametrization

To evaluate the impact of our proposed approach, we compare four different parametrizations for the predictions:

| Model | Prediction Method |
|---|---|
| **Baseline (Benchmark)** | Ordinary Kriging with only DWD stations. |
| **Naïve Fusion** | Ordinary Kriging with all weather stations. |
| **A-Priori Information** | Adapted GPR with a-priori values for uncertainty for all stations. |
| **Learned Model** | Proposed, new model. |

The baseline model represents the state-of-the art prediction without the benefit of VGI data. The Naïve Fusion represents the blind use of the additional data without any regard to the data quality. As to our knowledge this blind use has not been done before and shows the potential risks of VGI as well as a second benchmark for the use quality assessment. A-Priori Information represents the knowledge of experts regarding the quality of measurements, e.g. experience of prolonged use or specifications of sensors. In this work we determined the quality value for each station class by a simple grid-search. We assume in this parametrization that the quality of each sensor class is the same and do not differentiate between each single sensor. The DWD stations have a quality value of 0.98, the Wunderground sensors have a quality value of 0.81. The Learned Model represents the proposed new model. Bases on the A-Priori Information parameter, for each sensor a unique uncertainty value is learned iteratively with the presented combined model.

## Results and Discussion

The results for the temperature prediction can be seen in Table 1 and Fig. 3 graphically.

| MODEL | MEAN ABSOLUTE ERROR | STANDARD DEVIATION |
|---|---|---|
| **Baseline (Benchmark)** | 1.12°C | 0.83°C |
| **Naïve Fusion** | 1.26°C (-12.5%) | 1.03°C |
| **A-Priori Information** | 1.21°C (-8.0%) | 0.99°C |
| **Learned Model** | 0.98°C (12.5%) | 0.76°C |

*Table 1: Summary of prediction results in degree Celsius. In brackets the percentage improvement compared to the Benchmark.*

We use the Mean Absolute Error (MAE) as error metric as this shows the quality of the prediction in a single value and is well-established. The standard deviation (SD) is used to show the volatility of the quality of the results.

We see for all models that the ranking of the MAE and SD is the same for every model. Not surprisingly the naïve fusion model performs the worst. Without any quality assessment, the influence of false measurements and high variance in placement decreases the quality of the prediction compared to the traditional approach, the baseline model. The increased availability of information inherent in VGI is overshadowed by the poor and heterogeneous quality of the measurements. The inclusion of a very simple quality assessment with the a-priori information model shows already an increase in prediction quality. Even though there is no differentiation between the stations within each class, the quality increases even with this very simple approach. But it still performs worse than the baseline model. The learned model performs the best overall. Compared to the naïve fusion it performs more than 20% better. This is the result of the learning process and the used covariance function. Sensors which perform bad overall gain less weight to the prediction result over time. Low quality sensors are automatically filtered based on their data, e.g. when they are inside buildings or are defective and produce constant values.

But while the accuracy of the prediction is important, the increased spatial resolution of the prediction is one of main advantages VGI presents. Figure 3 shows the resulting predictions of the different models; the a-priori information model was left out as it is almost identical to the naïve fusion model.

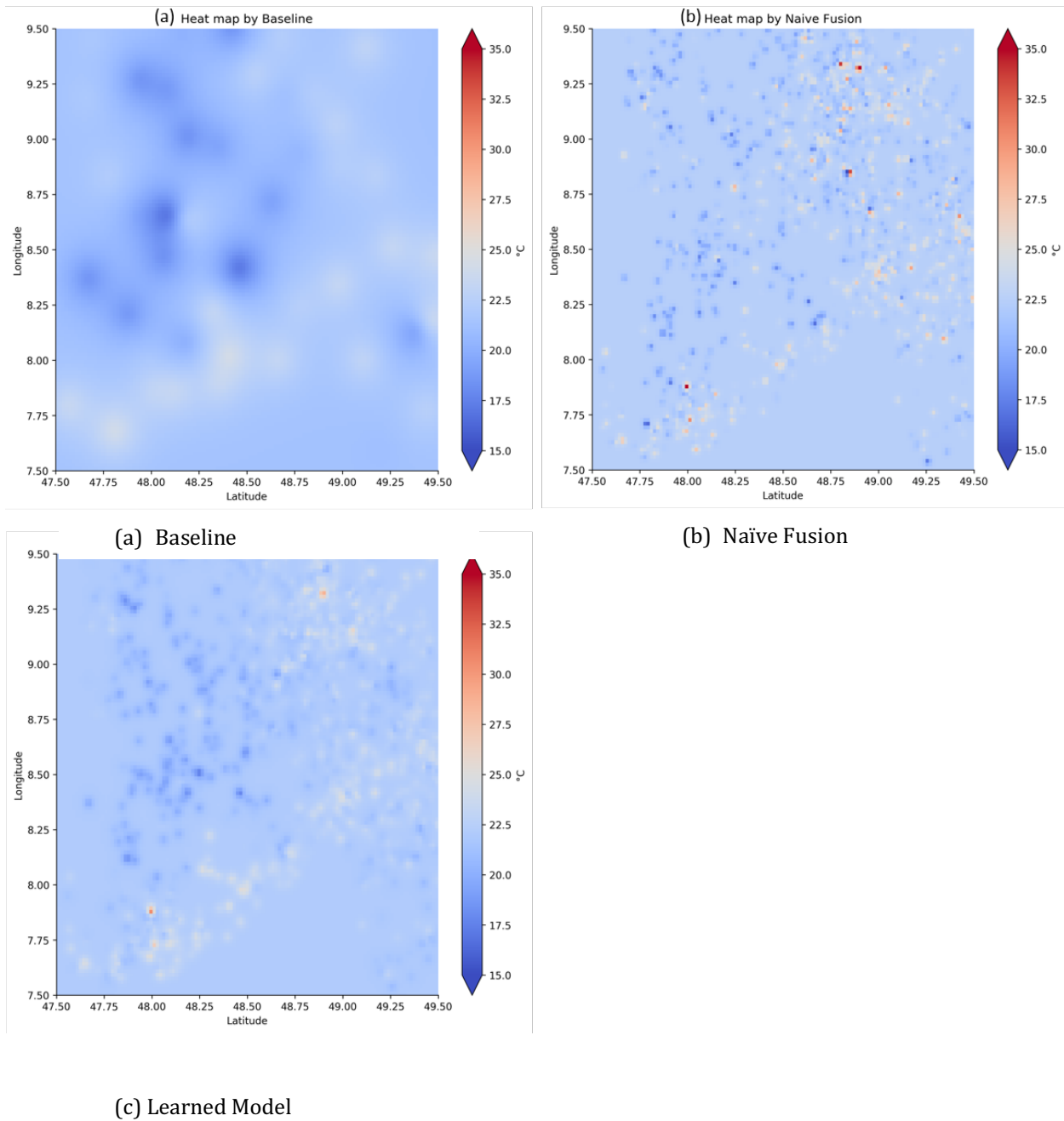

(c) Learned Model

*Figure 3: Resolution of predictions.*

We can see the differences in resolution and prediction for each model. The naïve fusion presents a highly detailed map of temperature as the number of available sensors is quite high. There is a strong, but fluid transition between the different areas. When compared to Fig. 1 the number of weather stations are directly connected to the sharpness of the map, which can be seen in the south and the west. Outliers are seen on this map. The temperature span is from 14.6°C up to 35.23°C. The baseline model shows a low spatial resolution, low overall temperature as well as a low span of the temperature, which ranges only from 17.27°C to 24.12°C. The low number of DWD weather stations can be seen by the rough transition between the different prediction areas. In the south-west of the map, there is an increased number of DWD stations and the map is much smoother. The learned model strikes a balance between the other models. The temperature is between 17.77°C and 27.52°C and the transitions are overall smooth. A clear distinction between warmer and colder areas can be seen and allow a detailed temperature map. The overall trend of the temperature distribution over our study area stays the same for all predictions. The graphical analysis shows clearly the advantages of VGI. Where in the pure prediction results the baseline outperforms the naïve fusion quite strongly, in the practical use to create informative maps, the naïve fusion outperforms in our opinion the baseline.

## Error Distribution

To further evaluate and understand the different models, we examine their error distribution. Fig. 4 shows the histogram of the prediction errors for each model. One can see that the naïve fusion as well as a-priori model are almost identical. This is not surprisingly, as these are quite similar in their parametrization and underlying modelling logic. Both resemble a broad normal distribution with an overall great span of around 10°C. Overall, there does exists a slight negative bias of the predictions, indicating that these models overestimate the temperature. This is most likely the result of the difference in placement of the reference stations and the citizen weather stations. As discussed before, the standardized placement of DWD stations leads to an exclusion of several climatic conditions, e.g. urban heat islands. These are captured by the citizen weather stations and lead to an overestimation of the temperature, as these effects are not filtered out. The baseline model on the other hand underestimates the temperature as the majority of its errors lies in the interval of 0 and 2, which lead to a skewed distribution. We assume this is the result of few outlier stations, which decrease the mean temperature of the overall distribution, in particular as stations in the warmest cities of Germany are in this area (Freiburg and Karlsruhe) as well as the different climatic regions such as the Black Forest and the Upper Rhine valley. But we also see the effect of the standardized placement as the standard deviation of the errors is lower than for the other two errors. This presents a more coherent prediction, which can also be seen when comparing the baseline to the naïve fusion in Fig. 3. Finally, the learned model shows a similar distribution as the baseline model, but the width of the distribution is even smaller and the center of the errors lies between 0 and 1. The shape is, similar to the graphical prediction in Fig. 3, a combination of the baseline and the naïve fusion models. The histogram supports the hypothesis that our approach in form of the learned model manages to leverage the advantages of VGI successfully.

Of further interest is that the highest errors for all models are negative. Interestingly, the outlier is the strongest for the baseline model. We would have expected that such a high error would only be present in VGI measurements. This indicates that there does exists at least one station in the reference stations used for the evaluation which has a relatively low temperature compared to all other stations nearby. Therefore even when using official data sets we would emphasize caution and a rigorous data understanding before applying these data sets for analysis and generation of insights.

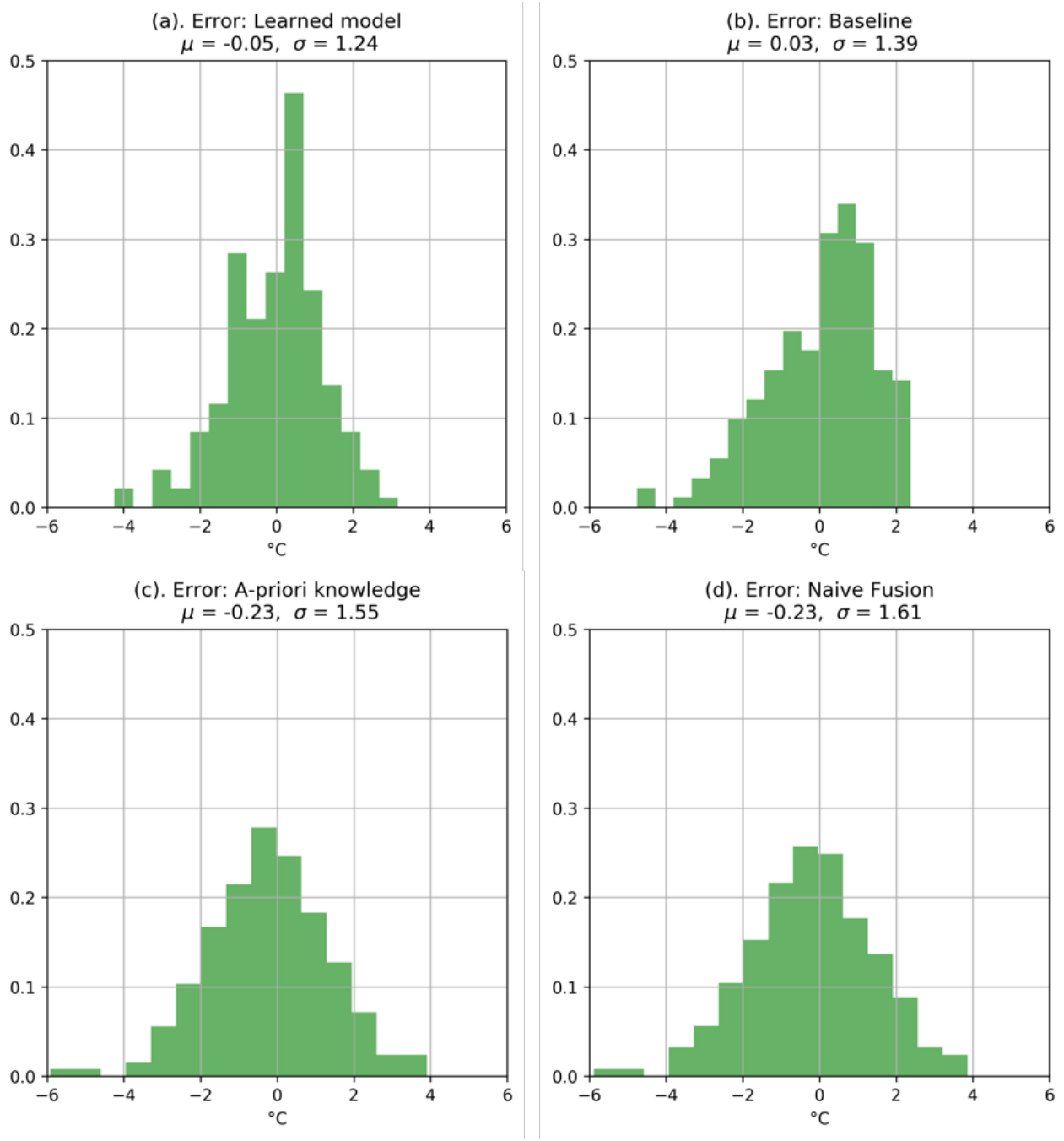

*Figure 4: Histogram of the Prediction Error Out Of Sample.*

## Conclusions and Future Work

In this work, we proposed an automated quality assessment of VGI sensors with weather stations as the concrete use-case. The proposed approach combined a new evolutionary learning algorithm with Gaussian Process Regression to learn and model the quality of sensors to produce reliable and accurate predictions without the need to clean the data beforehand. We evaluated the approach on weather data as this data provides the most accessible data and is therefore most likely to be used by researchers and practitioners alike. Our results showed an improvement in the prediction quality of 12.5% in comparison to the established benchmark of DWD weather stations, by only additionally including the quality of the measurement. Furthermore, we showed that, without regard to the prediction quality, the naïve use of citizen weather stations improves the spatial resolution of the temperature prediction immensely. The proposed approach preserved this improvement of the spatial resolution and managed to provide the full benefit of VG as discussed e.g. in Blaschke et al. (2011) and Meier et al. (2017). In particular for future smart cities and urban climates this allows for more in-depth analyses as to this date the existing measurement networks are rather sparse, e.g. for temperature and air pollutants, and new (crowdsourced) measurement approaches are developed and implemented such as in the SMARTAQNET project[4], or with cars as sensors. Our approach allows to fully benefit from those innovations.

This research has several restrictions which have to be taken into account. First, we evaluated our approach only on temperature data in South-West Germany in a short time-interval at the beginning of August 2016. While the overall data size for the training as well as evaluation data set is quite big, it is only based on a small fraction of the overall available data. In particular seasonal and daily cycles have not been examined. Second, we only fully implemented and compared one prediction method, the ordinary kriging, and one kernel approach to incorporate the uncertainty. While the reasoning for this is discussed in our method chapter, a more in-depth comparison could lead to different results. Third, we did not compare our results to a manually cleaned data set as in Meier et al. (2017). We assume this could lead to an improvement of the naïve as well as a-priori method, but this is by design beyond the scope of this work.

In the future, an evaluation with different data set would be of high interest, especially with air pollutants and in different climatic regions. Another interesting question would be the inclusion of different kernels as well as background information. The work of Bhattacharjee et al. (2016) shows an example with semantic kriging, which includes land use information and could be used as an alternative kernel to ordinary kriging. Another approach is found in regression-kriging, discussed in Hengl et al. (2007). Arnfield (2003) presents an overview of causal factors for the influence on temperature. The use of spatio-temporal prediction instead of only spatial prediction could lead to further insights. Kilibarda et al. (2014) show the application of such a spatio-temporal kriging and the benefits it provides. The challenge here lies in the selection and modelling of the suitable kernel as well as the computational complexity. Finally, the results of our error analysis of the baseline model show a strong skewness. Further investigations into this error could lead to interesting insights.

## Acknowledgements

Blinded for Review

---

[4] http://www.smartaq.net/